# An Industrial Case Study on Measuring the Quality of the Requirements Scoping Process

Krzysztof Wnuk

Software Engineering Research Lab, Department of Software Engineering Blekinge Institute of Technology Karlskrona, Sweden

`Krzysztof.Wnuk@bth.se`

Markus Borg

SICS Swedish ICT AB, Lund, Sweden

`Markus.Borg@sics.se`

Sardar Muhammad Sulaman
Department of Computer Science, Lund University, Sweden
`sardar@cs.lth.se`

Decision making and requirements scoping occupy central roles in helping to develop products that are demanded by the customers and ensuring company strategies are accurately realized in product scope. Many companies experience continuous and frequent scope changes and fluctuations but struggle to measure the phenomena and correlate the measurement to the quality of the requirements process. We present the results from an exploratory interview study among 22 participants working with requirements management processes at a large company that develops embedded systems for a global market. Our respondents shared their opinions about the current set of requirements management process metrics as well as what additional metrics they envisioned as useful. We present a set of metrics that describe the quality of the requirements scoping process. The findings provide practical insights that can be used as input when introducing new measurement programs for requirements management and decision making.

## 1  Introduction

Requirements Management (RM) [4] iteratively integrates the requirements elicitation and analysis results into the project management and development flows. RM also supports managing requirements during the product lifecycle and between the products. Large, globally operating software companies need to manage large quantities of features and requirements that continuously arrive from ever-changing markets [10].

Measuring and optimizing requirements identification, prioritization, definition and implementation processes is, in a market-driven context [10], crucial for achieving and sustaining competitive product growth [5].

The process of selecting a subset of requirements for implementation within a given project is called scoping. Many software-intensive companies increase the flexibility of decision making by allowing scope fluctuations. Our previous work highlighted that large companies experience frequent scope fluctuations and have limited support in scope management [14]. The resulting late changes increase the need for improved monitoring and management capabilities that can evaluate the adequacy of the selected requirements management process models. Despite that, most published work on requirements measurement focus on the attributes of requirements [2] rather than the requirements management process [1]. Some published process metrics include: i) how much value a software team delivers in every iteration [3], ii) the number of requirements awaiting analysis, prioritization or decision, and iii) the lead time in each state for each user story [7].

In this paper, we present the results from an exploratory interview-based case study among 22 participants working with the requirements management process at a large company that develops embedded systems for a global market. During semi-structured interviews with mostly senior-level practitioners working with requirements gathering, prioritization, scope management, software resource planning and high-level management, we investigated the following two research questions:

**RQ1:** What are the current scope management process quality metrics used by the case company?

**RQ2:** What scope management process quality metrics would the practitioners like to implement in the future?

## 2  Case Company

The case company is a large (5,000 employees) organization active in the telecommunication domain, developing embedded systems for the global consumer market. As the inflow of new requirements is rapid, product management often needs to make unplanned scoping decisions [14]. The company utilizes the Software Product Lines concept [9]where different development projects contribute to an evolving common code base, also called a platform. The total number of features registered in the company's database exceeds 10,000 and is steadily growing as new products are added to the product line, each containing on average 60 to 80 new features and associate up to 20 system requirements per feature. Feature implementation is allocated to approximately 20 to 25 development teams (each team has from 40 to 80 developers).

Features are managed based on a state machine depicted in Figure 1. When features are created, they are put into an administrative state called New Feature (NF). In the next step, features enter the process and are discussed at the M0 forum. This forum critically reviews if a proposed feature has a sponsor, sufficient business justification and is aligned with the current product and portfolio strategy. Many features are rejected at this stage mainly due to insufficient business justification or unclear defini-tion. Next, a feature is promoted to the M1 state where it is prioritized against other features by scope owners using a one-dimensional prioritization based on business value. A feature could be returned to the M0 state for further refinement.

At the next stage, called M2 in Figure 1, the development resources are consulted and implementation schedules are discussed and agreed upon. Each feature comes to this forum with a target delivery date that is discussed and adjusted depending on the current software development organization load and other responsibilities. Prototypes are used at this stage to provide more accurate effort estimates and possible delivery times. A pipeline tool is used at the M2 stage to control the resources and to schedule delivery of several hundreds of implementation running in parallel. After the M2 state, the development organization takes the main responsibility for the features which are promoted to the: Definition Ongoing (DO), Awaiting Execution (AE), Execution Started (ES), Execution Completed (EC), Awaiting Integration (AI) and Integrated (I). Transitions between any two states are in theory allowed, including backward transitions. However, there is one optimal path without backward transitions through a state machine which is visualized with a dashed line in Figure 1.

## 3      Research Methodology

To gain deep understanding and explore different requirements management metrics at the case company, a flexible case study design was chosen [11] with semi-structured interviews as the method for data collection.

We started phase 1 by iteratively developing an interview instrument in collaboration with four practitioners from the case company. Finally, two senior software engineering researchers and two practitioners reviewed the 17 questions and they were grouped into 6 topics: background, business goals, current metrics, desired metrics, visualization, and open innovation. The interview instrument can be accessed online [12]. Note that the results reported in this paper are exclusively related to questions under topics 3 and 4 as well as background questions under topic 1. The remaining topics are covered in a separate publication [15].

In the next step, we selected interview respondents by using a combination of max-imum variance and convenience sampling [11] to cover as many views on the requirements management process as possible. Twenty-two respondents participated in the study. Their average experience in working with requirements processes was 6,5 years with the most experienced participant having almost 13 years of experience and the least experienced participant having about 3 years of experience [12].

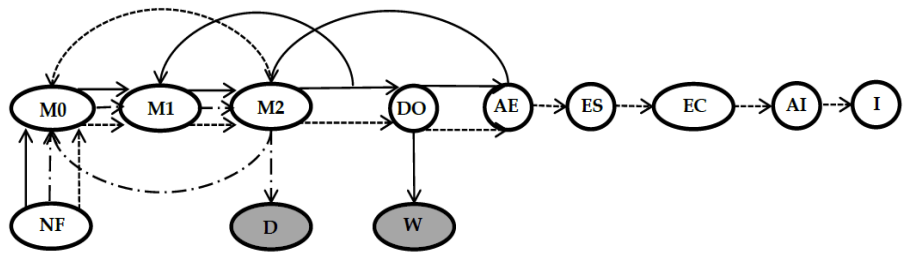

**Fig 1. The example history of three features. The first feature, marked with dashed lines was implemented. The second feature marked with solid arrows was withdrawn. The third feature marked with dashed-dotted arrows was discarded. Also available at [13] .**

Prior to the interviews, we sent the questions to the participants to help them understand the scope of the study and prepare for the discussions under the interviews. The first author then interviewed all participants individually, recorded and transcribed the interviews. The transcripts were sent to the interviewees to validate the content, and to enable clarifications where needed.

As the industry partner requested a quick summary of the findings, we concluded the first phase with five senior managers in a seminar. During this seminar, the first author presented preliminary results from an initial analysis of the data. The seminar delivered early tangible outcomes and the discussions at the seminar also acted as a validation, i.e. a sanity check that the direction of our work was promising, and motivated the deeper analysis in phase 2.

Phase 2 involves the four steps of the systematic data collection and analysis. First, the first author divided the transcripts into chunks of text containing a few connected sentences. The second author then repeated the process for 4 of the interviews (21%), to validated that we had a reasonable level of granularity. The authors compared the chunk sizes, and agreed on simple rules resulting in the creation of a chunk for each relevant proposition (i.e. what is believed, doubted, etc.) expressed in the interviews. The first author then reiterated the remaining chunks to apply the rules.

In the second step, the first and second authors collaboratively analyzed 11 of the interviews (48%) with the goal of developing a robust coding scheme. The first and second authors then independently coded the remaining 12 interviews (52%). The authors calculated an inter-rater agreement using Cohen's Kappa [6] on the coding results. We achieved a Kappa score of 0.59, which we interpret as moderate agreement on the coding scheme.

In the third step of phase 2, we analyzed the coded data. The output from the coding step was synthesized by the first and second authors, and reviewed by the third author to provide further validation and observer triangulation. Finally, all authors prepared the manuscript for this research article.

**Validity.** We discuss validity issues based on the guidelines by Runeson et al. [11]. We attempted to mitigate the *interpretive validity* threats by asking interviewees to check the interview transcripts. Threats to *evaluative validity* are not applicable in this case due to exploratory nature of the study and a lack of evaluative purpose. Threats to *description validity* were addressed by recording the interview sessions and transcribing them. The transcripts were sent back to the interviewees for validation. Threats to *theoretical validity* have a minimal impact on this study due to its explorative nature and therefore a lack of theory, specific hypotheses, or conceptual frameworks to be validated. Moreover, we minimized the bias of unclear questions by iteratively developing the interview guidelines. The questions were formulated in a way to minimize the possibility of imposing a particular answer. We took precautions that the interviewer expressed neutrality when asking the questions and therefore the risks of reflexivity are minimized.

Due to an exploratory nature of this study, exploring to what extent our conceptualizations and conclusions derived from the interviews are correct remains to a certain degree unclear and calls for inspection by other researchers in the field as well as follow-up studies. Since the investigated problem originates from the case company, we can for sure claim that it is an authentic research problem.

We report that both *internal and external generalizability* are strongly limited in this case, mainly due to only one company involved. The paper's exclusive focus on an individual company narrows the applicability of the observations. Nevertheless, we attempted to gather as many perspectives as possible on the studied phenomenon by inviting participants with various roles and experiences from the case company.

## 4. Results

Table 1 summarizes the 26 scope management process quality metrics identified in the study. Among them, only five metrics are measured and 21 are needed or requested metrics. The five currently used metrics are: the number of backward transitions (Q1) and their reasons (Q2), the software design quality and if the process actually prioritizes the most important features from the portfolio planning (Q4) and customer perspectives (Q5).

The requested metrics include the impact of priorities on the lead-times (Q6 and Q7) as well as the impact of high priority features on low priority features (Q8). The accuracy of estimates and its impact on the efficiency of requirements analysis or definitions (Q9 and Q12) clearly indicate that focusing plainly on speed may not give the desired effects as quality of the work should not be compromised.

Several metrics also describe the features and their nature in terms of testability or complexity, e.g. Q10 and Q11. These metrics should be introduced during the requirements analysis phase and used as extensions to the widely accepted aspects, e.g. correctness, ambiguity or completeness. Q10 and Q11 further detail requirements on system test metrics suggested by Petersen and Wohlin [8].

Metrics Q14 and Q15 focus on how many times or why a feature was sent back in the process due to unclear information. This indicates that some stages of the process may either not do their work rigidly or receive appropriate input from earlier stages - thus delivering requirements of insufficient quality.

Similarly, our respondents would like to measure how many times a feature is moved between the releases (Q13) and why they were moved, which could indicate either: i) issues with accurate release planning or ii) several strategic changes after the release plan is agreed upon. Metric Q13 provides interesting input for the iterative release planning approaches that are based on continuous release re-planning and timely responses to a frequently changing market situation. The number of release changes could be correlated with how many times previously set delivery dates are altered (Q25).

Two requested metrics focus on the "waste" generated by analyzing unimplemented features (Q18 and Q19) while one metric (Q20) focuses on the effort saved on unimplemented features or software definitions in relation to the previous (waterfall) way of working. Two other requested metrics correlate the defined scope with the overall product strategy (Q24) or increased sales from successful products (Q23).

Finally, the amount of changes in the open source code each feature requires (Q22) in combination with the percentage of effort put on legacy work while developing new features (Q26) can bring interesting insights regarding the selected sourcing strategy and also suggestions about the amount of open source code in the product.

**Table 1.** Elicited metrics. Respondents are coded with Greek alphabet letters.

| ID | Metric Definition | Mentioned or need for/ Respondent |
|---|---|---|
| Q1 | How many times a feature is sent backwards in the process | Measured, ZETA |
| Q2 | Why features are sent backwards in the process from the M2 forum | Measured, RHO |
| Q3 | The quality of the software design and the associated user interaction features | Measured, TAU |
| Q4 | The overlap or potential discrepancies between the early product definitions from the portfolio planning and the product scope at the TG Commit | Measured, SIGMA, ETA |
| Q5 | Priority levels of highly requested features by various stakeholders. | Measured, KAPPA |
| Q6 | The correlation between the priorities set and the time needed to implement the features or the time needed for definition or implementation. | Need for, GAMMA, LAMBDA |
| Q7 | The frequency of priority changes in relation to the dev. performance | Need for, PI |
| Q8 | How new-coming highly-prioritized features impact low-priority features (e.g. low-prioritized get delayed) | Need for, RHO |
| Q9 | The accuracy of the estimates in relation to the efficiency of the process. | Need for, ZETA |
| Q10 | The testability of the "vertical features" (features involving several technical areas) and their impact on various technical areas. | Need for, THETA |
| Q11 | The complexity of the features that are sent to the definition (cf. DO in Figure 1) in terms of their impact on other organizations. | Need for, THETA |
| Q12 | The quality of feature definitions and estimates. | Need for, KAPPA |
| Q13 | How many times features are moved between the releases and why. | Need for: BETA, ETA, KAPPA, MY |
| Q14 | How many times (and why) the features are send back from the M1 to the M0 forum. | Need for ALFA and CHI |
| Q15 | How many times (and why respondent CHI) features are send back from M2 to M1 forum (respondents JOTA and LAMBDA), | Need for, CHI, JOTA, LAMBDA |
| Q16 | How many time a feature is resubmitted at the M0 forum due to unclear information or quality issues | Need for, LAMBDA |
| Q17 | The reasons why features are send back from the M0 forum to redefinitions | Need for, RHO |
| Q18 | How much "waste" the process is producing (analyzed but unimplement- | Need for, PHI, CHI |

| | ed features, e.g. how many features are withdrawn at each stage). | |
|---|---|---|
| Q19 | The "waste of the scope" after the features are promoted to the definition. | Need for, KSI |
| Q20 | The effort saved on unimplemented features or software definitions in relation to the previous (waterfall) way of working. | Need for, PHI |
| Q21 | The stability of the scope after the customer acceptance test. | Need for, KAPPA, NY |
| Q22 | How many changes to the OSS code each feature requires and when to share these changes with the open source community. | Need for: PI |
| Q23 | If the planned scope later implemented in the products is meeting the set sales and customer satisfaction business targets | *Need for:* ETA |
| Q24 | To what degree the features that are created in the process reflect the overall strategy of the company. | Need for: PI |
| Q25 | How many times previously set delivery dates altered (caused by e.g. resource shortages or changed priorities). | *Need for:* ETA |
| Q26 | The percentage of effort put on legacy work | Need for: CHI |

## 6. Implications and Conclusions

Our study delivers several implications for research and practice. Firstly, the fact that we elicited 26 quality metrics is a clear indication that it is challenging to come up with an accurate set of metrics to capture the important aspects the requirements management process. Secondly, the fact that 21 requested quality metrics ideas were collected brings a possible interpretation that more focus should be directed towards complementing efficiency metrics with quality metrics. For example, quickly delivering features with minimal process waste is highly desired, as long as these features will provide value to the end customers and positively realize product strategies, see for example metrics Q23 and Q24.

Thirdly, requirements prioritization for agile development should go beyond popular one-dimensional priority or urgency lists and be correlated with measures that take the holistic perspective on prioritization (see e.g. metrics Q5, Q6 and Q8) and integrate it with product and portfolio planning.

Fourthly, measuring lead-times and delays on the interface between the requirements and development organizations appears to be equally important as measuring the requirements process lead-times. Additional significant factor is to measure backward transitions and understand why they happen (see metrics Q1, Q2, Q14, Q16 and Q17) or transitions between the releases (Q13). Fifthly, measuring the number of features in each state should be complemented with the derived measures of the ratios between the features in two states. This provides useful indications for rapid identification of process bottlenecks.

In future work, we plan to create a conceptual model of measuring and tracking potential waste in requirements management and decision making processes. Moreover, we plan to conduct additional case studies at other companies that record the information during their requirements management processes. Such empirical studies would help us in expanding our knowledge about the applicability of our model.

**Acknowledgements.** This work is supported by the IKNOWDM project from the Knowledge Foundation in Sweden (20150033).